\documentclass[aps,preprint,showpacs]{revtex4-1}
\usepackage{graphicx,subfigure}
\usepackage{amsmath,amsfonts,amssymb}
\usepackage[hypertex]{hyperref}

\newcommand{\f}{\frac}
\newcommand{\suml}{\sum\limits}

\graphicspath{{D:/localtexmf/Mytex/LGH/mathplots/}}
\begin{document}
\title[Global isomorphism]{The application of the global isomorphism to the study of liquid-vapor equilibrium in
two and three dimensional Lenard-Jones fluids}
\author{V.L. Kulinskii}
\email{kulinskij@onu.edu.ua}
\affiliation{Department for Theoretical
Physics, Odessa National University, Dvoryanskaya 2, 65026 Odessa, Ukraine}
\begin{abstract}
We analyze the interrelation between the coexistence curve of the Lennard-Jones fluid and the Ising model in two and three dimensions within the global isomorphism approach proposed earlier [V.~L. Kulinskii, J. Phys. Chem. B \textbf{114} 2852 (2010)]. In case of two dimensions we use the exact Onsager result to construct the binodal of the corresponding Lennard-Jones fluid and compare it with the results of the simulations. In the three dimensional case we use available numerical results for the Ising model for the corresponding mapping. The possibility to observe the singularity of the binodal diameter is discussed.
\end{abstract}
 \maketitle
The connection between continuous and discrete systems in
Statistical Physics is the source of new methods and fruitful
applications. The lattice models are usually considered as the
caricatural pictures of the real ones. Due to their numerical and
analytical tractability they serve as the main sources of the
rigorous results. Famous Onsager's solution of the two-dimensional
Ising model is the well known example which gave new impact for the
theory of Critical Phenomena. In the modern theory of the Critical
Phenomena the ideology of the isomorphism classes of the critical
behavior provides the description of the real systems using the
results obtained for the model systems among which the lattice models play important role \cite{book_baxterexact}.
In particular the molecular liquids with
short range interactions of the Lenard-Jones (LJ) type:
\begin{equation}\label{ljpotential}
  \Phi(r) =
  4U_0\left(\,-\left(\,\f{\sigma }{r}\,\right)^6 + \left(\,\f{\sigma }{r}\,\right)^{12}\,\right)\,,
\end{equation}
belong to the isomorphism class of the Ising model or equivalently
to the Lattice Gas (LG) model. The latter is determined by the
Hamiltonian:
\begin{equation}\label{ham_latticegas}
  H = -J\suml_{
\left\langle\, ij \,\right\rangle
  } \, n_{i}\,n_{j} - \mu \,\suml_{i}\,n_{i}\,,
\end{equation}
where $n_{i} = 0,1$ whether the site is empty or occupied
correspondingly. The quantity $J$ is the energy of the site-site interaction of the nearest sites $i$ and $j$,
$\mu$ is the chemical potential.
The order parameter is the probability of occupation
$x = \left\langle\, n_i\,\right\rangle$ and serves as the
analog of the density. The phase diagram of the LG is
symmetrical with respect to the line $x_{0} = 1/2$
and formally extends up to the region $T\to 0$,
where the limiting states $x=0$ and $x = 1$ exist only.

In \cite{eos_zenome_jphyschemb2010} the approach which extends the notion of the isomorphism between the LG and the LJ fluid from the critical region to the whole liquid-vapor part of the phase diagram was proposed. It is based on the fact that the line $x=1$ of
the LG can be thought of as the analog of the Zeno-line for liquid \cite{eos_zenoboyle_jphysc1983,eos_zenobenamotz_isrchemphysj1990}.
In construction of the mapping the results of works of Aphelbaum and Vorob'ev \cite{eos_zenoapfelbaum_jchempb2006,eos_zenoapfelbaum_jchempb2008}
on the Zeno-line was extensively used.

The objective of this paper is to use the global isomorphism
transformation to the construction of the binodal of the
Lennard-Jones fluids.
We relax the condition of the constancy for the compressibility factor used in above mentioned works of Aphelbaum and Vorob'ev and interchange less restrictive. We will demonstrate that in such case we are able to get rather good estimates for the locus of the critical points and map the binodals of the Ising model to that of the Lennard-Jones (LJ) fluid.


The following mapping representing the global isomorphism between the thermodynamic states of the LG and the LJ fluid was constructed:
\begin{equation}\label{projtransfr}
  n =\, n_b\,\f{x}{1+z \,t}\,,\quad
  T =\, T_Z\,\f{z\, t}{1+z \,t}\,,
\end{equation}
with
\[z = \f{T_c}{T_Z-T_c}\,.\]
Here $n$ and $T$ are the density and the temperature of the fluid, $t$ is the temperature variable of the LG normalized to the critical temperature so that $t_c = 1$ we also use the standard dimensionless values for $T$ and $n$ of the LJ fluid \cite{book_hansenmcdonald}.
$T_Z$ and $n_b$ are the parameters of the linear element:
\begin{equation}\label{vdw_z1}
  \f{n}{n_b}+\f{T}{T_Z} = 1\,.
\end{equation}
The coordinates of the CP for the fluid are:
\begin{equation}\label{cp_fluid}
  n_{c} = \f{n_B}{2\left(\,1+z\,\right)}\,,\quad   T_{c} = T_Z\, \f{z}{1+z}\,.
\end{equation}

In contrast to the approach of \cite{eos_zenoapfelbaum_jchempb2006} the parameters $T_Z$ and $n_b$ are chosen from the condition of consistency with the van der Waals (vdW) approximation for the given equation of state (EoS). In particular, $T_Z$ equals to the Boyle temperature in the vdW approximation.  The parameter $n_b$ is chosen from the condition of linear change
\begin{equation}\label{dzdt}
    \f{d}{dT} \,\f{Z(n(T),T)- 1}{n(T)} =  0\,,
\end{equation}
of the compressibility factor $Z$:
\begin{equation}\label{zpt}
  Z =\f{P}{n\,T} = 1-\f{2\pi \,n}{3\,T}\int r^3
\frac{\partial\, \Phi(r)}{\partial\, r}\,g_2(r;n,T)\, d\,r\,,
\end{equation}
along the linear element \eqref{vdw_z1}.
Using the virial expansion it is easy to get the condition:
\begin{equation}\label{nbtz}
n_b= T_Z  \,\f{B'_2\left(\,T_Z\,\right)}
{B_3\left(\,T_Z\,\right)}\,,
\end{equation}
which formally coincides with the standard one (see e.g. \cite{eos_zenobenamotz_isrchemphysj1990}). The difference is that we relax the condition of the constance $Z$ along the linear element \eqref{vdw_z1} since it is valid only for the vdW EoS. As was shown in \cite{eos_zenomeglobal_jcp2010} such definitions along with the condition
\begin{equation}\label{zds}
 1/z  =1+s/d\,,
\end{equation}
following from the scaling properties of the attractive potential $\Phi_{\text{attr}} \simeq r^{-(s+d)}$ give the estimates for the CP loci of $2D$ and $3D$ LJ fluids which are in good agreement with the results of the numerical simulations.

\begin{figure}
\center
  \includegraphics[scale=1]{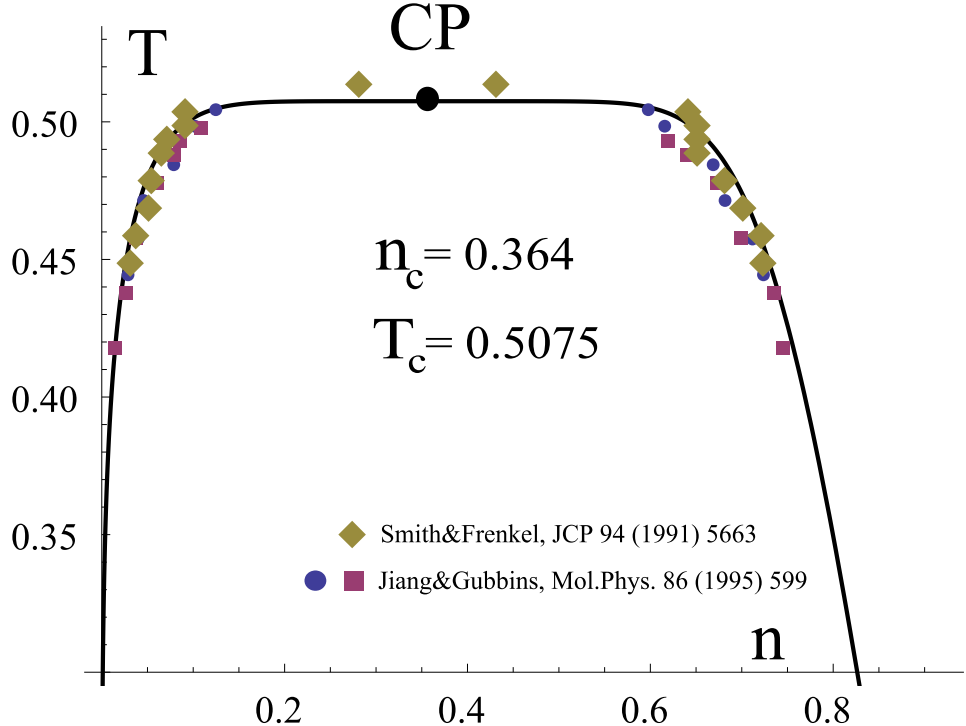}\\
  \caption{Fitting the binodal of two dimensional LJ fluid \eqref{projtransfr} with $z=1/3$, $T_Z = 2.03$ and $n_b = 0.971$ to the results of the simulations \cite{crit_lj2dim_jcp1990,crit_lj2d_molphys1995}. Also the corresponding locus of the CP is shown.}\label{fig_simulcomparis}
\end{figure}

To prove that such closeness is not a coincidence we check another consequence of \eqref{projtransfr}. The latter provides the mapping between the binodals of the LG (Ising model) and LJ fluid. In accordance with \eqref{zds} $z = 1/3$ for $2D$ LJ fluid and $z=1/2$ for $3D$ case respectively. For $2D$ case the known exact solution for $2D$ Ising model \cite{crit_onsager_pr1944} is:
\begin{equation}\label{isingbinodal2}
  x = 1/2\pm f(t)^{1/8}\,,\quad f(t) = 1-\f{1}{\sinh^4\left(\,2J/t \,\right)}\,.
\end{equation}
Substituting it into \eqref{projtransfr} the parametric representation for the binodal of LJ fluid is obtained and can be compared with the known numerical simulation results (see Fig.~\ref{fig_simulcomparis}). Note that the values of the parameters $T_Z$ and $n_b$ obtained by fitting are very close to that following from the consideration above $T_Z \approx T_{B}^{vdW} = 2$ and $n_b \approx 0.91$  (see also \cite{eos_zenomeglobal_jcp2010}). The locus of the CP obtained is in good correspondence with the results of \cite{crit_lj2d_molphys1995}.

The same procedure can be applied to the $3D$ case. Here we use the results of \cite{crit_3disingmc_jmathphys1996} for the $3D$ Ising model. There data for the spontaneous magnetization $M(t) = 2x(t)-1$ were described by the expression:
\begin{equation}\label{eos_3dising_blote}
 x =  \frac{1}{2} \left(1\pm\tau^{\beta} \left(b_0+a_1 \tau+
   a_{\Delta_1}\,\tau^{\Delta_1}\right)\right)\,,\quad \tau = 1-t\,,
\end{equation}
where $\beta= 0.326941\,,\Delta_1 = 0.50842$, $ b_0 = 1.6919\,, a_1 = -0.42572366\,, a_{\Delta_1} =  -0.34357731$. This expression represents the data in the region $0.0005<\tau<0.26$. So we use \eqref{eos_3dising_blote} as the ``true`` binodal of the $3D$ LG. From the other side for the real molecular fluids the Guggenheim's \cite{pcs_guggenheim_jcp1945} empirical expressions:
\begin{equation}\label{guggenheim_binodal}
\f{n_{l,\,g}}{n_c} = 1+\f{3}{4}\,\left(\,1-T/T_c \,\right) \pm \f{7}{4}\,\left(\,1-T/T_c \,\right)^{1/3}\,.
\end{equation}
are well known.

\begin{figure}
\center
  \includegraphics[scale=1]{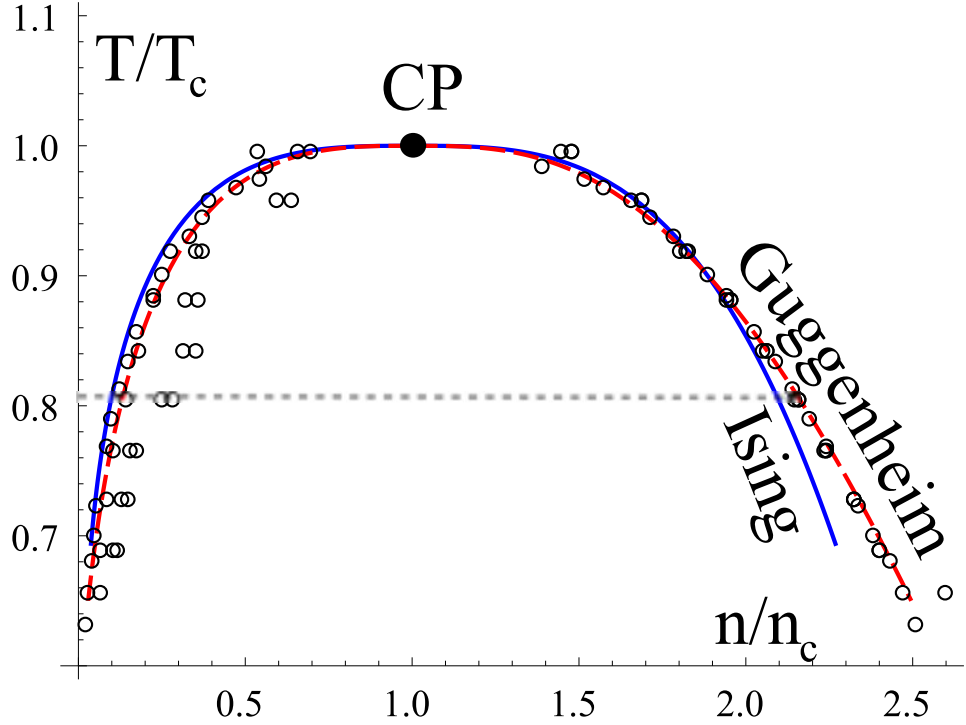}\\
  \caption{Binodal for $3D$ LJ fluid (blue line) obtained from \eqref{eos_3dising_blote}  the mapping after transformation \eqref{projtransfr} with $z=1/2$, $T_Z = T_B^{(vdW)} = 4$. The grey dashed line shows the range of applicability of \eqref{eos_3dising_blote}. The red line is for the Guggenheim binodal \eqref{guggenheim_binodal}. Circles are for the data \cite{eos_ljpcs_jcp2001,eos_ljfluid_jcp2005} (different circles at a given temperature in the gas branch are for different total number of particles in simulations of \cite{eos_ljpcs_jcp2001}).}\label{fig_lj_ising_guggenheim}
\end{figure}

In Fig.~\ref{fig_lj_ising_guggenheim} the result of transformation of the Ising binodal \eqref{eos_3dising_blote} is shown. The mapping of the Ising model quite satisfactory correlates with the data near the CP ($\tau <0.1$), while the Guggenheim curve better reproduce the data in outer region far from the CP.

Note that both \eqref{eos_3dising_blote} and \eqref{guggenheim_binodal} do not account for the diameter singularity of the density \cite{crit_buckham_ptcp1972,*crit_diamgreermold_annrevpc1981,*crit_diamexp_annrevchemphys1986}. The possibility to apply the transformation \eqref{projtransfr} directly to the results of simulations of LJ fluids is facilitated by the fact that it neglects the singular fluctuational corrections to the regular behavior of the diameter. According to the theoretical approaches \cite{crit_rehrmermin_pra1973,crit_fishmixdiam1_pre2003,crit_can_diamsing_kulimalo_physa2009} the diameter of the density of the molecular liquid has the following structure:
\begin{equation}\label{nd}
n_d = \f{n_l+n_g}{2\,n_c}-1 = D_{2\beta}\,|\tau|^{2\beta} + D_{1-\alpha}\,|\tau|^{1-\alpha} + D_{1}\,|\tau| + \ldots
\end{equation}

To the best of our knowledge none of the leading singular terms $\tau^{\beta}$ or $\tau^{1-\alpha}$ has never been discussed in computer simulations. According to \cite{crit_liqvamiepotent_jcp2000} the diameter anomaly is difficult to observe
by molecular simulations.
The data of computer simulations \cite{crit_lj2dim_jcp1991,eos_lj_molphys1992,crit_ljfluid_pre1995,
crit_liqvamiepotent_jcp2000,crit_longrangecampatey_jcp2001,eos_ljfluid_jcp2005,eos_yukawa_jcp2007} are claimed to be consistent with the law of rectilinear diameter and commonly, the locus of the CP is obtained by the extrapolation of the data from the two-phase region using the classical rectilinear diameter law. Then the value $n_c$ is greatly influenced by the form of such extrapolation in view of wide opening of the binodal.

It seems evident that LJ fluid does not possess any particle-hole symmetry which lead to regular diameter
and the density variable should display both anomalies in its diameter as it does in the case of real and model fluids \cite{crit_2beta_jcp1983,*liqmetals_singdiamhensel_prl1985,*crit_aniswangasymmetry_pre2007}.
The results obtained for real substances show that $\left|D_{1-\alpha}/D_{2\beta}\right|\gtrsim 10$ \cite{crit_fisherdiam_chemphyslet2005,crit_aniswangasymmetry_pre2007}. So at least $|\tau|^{1-\alpha}$ anomaly may be observed.
This poses the problem as to the values of the critical amplitudes $D_{2\beta},\,D_{1-\alpha}$ and their dependence on the microscopic interactions. This question needs further investigation.

As a summary we have demonstrated the usefulness of the transformation \eqref{projtransfr} in order to get the binodal of LJ fluid. By this we also confirmed the adequacy of the construction of the linear element \eqref{vdw_z1} as the key point in construction of the isomorphism transformation \eqref{projtransfr}. This gives the alternative to the commonly used Zeno-line based considerations \cite{eos_zenobenamotz_isrchemphysj1990,eos_zenoapfelbaum_jchempb2006}.


%

\end{document}